\documentclass[12pt]{article}

\usepackage{layout}
\usepackage[textwidth=460pt, textheight=630pt,voffset=30pt]{geometry}

\usepackage{graphicx}%
\usepackage{multirow}%
\usepackage{amsmath,amssymb,amsfonts}%
\usepackage{amsthm}%
\usepackage{mathrsfs}%
\usepackage{mathtools}
\usepackage{stmaryrd} % for \llbrackets
\usepackage{accents} % for \llbrackets
\usepackage{tensor}

\usepackage[title]{appendix}%
\usepackage{xcolor}%
\usepackage{textcomp}%
\usepackage{manyfoot}%
\usepackage{booktabs}%
\usepackage{algorithm}%
\usepackage{algorithmicx}%
\usepackage{algpseudocode}%
\usepackage{listings}%
\usepackage{etoolbox} % For Definitions, theorems, ...
\usepackage{enumitem}
\setlist[enumerate,1]{label={(\roman*)}}

\usepackage{array}
\newcolumntype{P}[1]{>{\centering\arraybackslash}p{#1}} % For centering while using p in tabular

\usepackage[pdftex,colorlinks=true, pdfstartview=FitV, linkcolor= Mlink, citecolor=Mlink, urlcolor= Mlink, hyperindex=true, hyperfigures=true]{hyperref}
\usepackage{authblk}
 \usepackage[numbers,sort&compress]{natbib}
\IfFileExists{aastexv6.sty}{\input{aastexv6.sty}}{\input{/Users/quentinvigneron/Documents/Travail/Research/tex_/aastexv6.sty}}% For ADS abbreviations
\IfFileExists{Short_cut_Blind.sty}{\input{Short_cut_Blind.sty}}{\input{/Users/quentinvigneron/Documents/Travail/Research/tex_/Short_cut_Blind.sty}}

\newcommand{\rhoNew}{\delta\rho_\sigma}

\title{\bf Is expansion blind to the spatial curvature?}

\author[1]{{Quentin} {Vigneron}\footnote{\href{mailto:quentin.vigneron@umk.pl}{quentin.vigneron@umk.pl}}}
\author[2]{{Vivian} {Poulin}\footnote{\href{mailto:vivian.poulin@umontpellier.fr}{vivian.poulin@umontpellier.fr}}}

\affil[1]{\small\it
	{Institute of Astronomy, Faculty of Physics, Astronomy and Informatics}, 
	{Nicolaus Copernicus University}, 
	{{Grudziadzka 5}, {Toru\'n}, {87-100}, 
	{Poland}}}
\affil[2]{\small\it
	{Laboratoire Univers \& Particules de Montpellier (LUPM), CNRS \& Université de Montpellier (UMR-5299)}, 
	{{Place Eugène Bataillon}, 
	{F-34095 Montpellier Cedex 05}, 
	{France}}}

\begin{document}
%\layout

\maketitle


\begin{abstract}In [\href{https://arxiv.org/abs/2309.10034}{arXiv:2309.10034}], we proposed and motivated a modification of the Einstein equation as a function of the topology of the Universe in the form of a {bi-connection} theory. The new equation features an additional ``topological term'' related to a second non-dynamical reference connection and chosen as a function of the spacetime topology. In the present paper, we analyse the consequences for cosmology of this modification. First, we show that expansion becomes blind to the spatial curvature in this new theory, i.e. the expansion laws do not feature the spatial curvature parameter anymore (i.e. $\Omega_{\not= K} = 1, \ \forall \, \Omega_K$), while this curvature is still present in the evaluation of distances.
Second, we derive the first order perturbations of this homogeneous solution. Two additional gauge invariant variables coming from the reference connection are present compared with general relativity: a scalar and a vector mode, both sourced by the shear of the cosmic fluid.
Finally, we confront this model with observations. The differences with the $\Lambda$CDM model are negligible, in particular, the Hubble and curvature tensions are still present. Nevertheless, since the main difference between the two models is the influence of the background spatial curvature on the dynamics, an increased precision on the measure of that parameter might allow us to observationally distinguish them.

\end{abstract}

\newpage

\section{Introduction}

In Ref.~\citep{2022_Vigneron_c}, we showed that the non-relativistic limit of the Einstein equation is only possible if the spatial topology is Euclidean, i.e. for which the covering space is $\mE^3$. %(we direct the reader to Section~2.1 of \citep{2022_Vigneron_c} for a precise definition of this term). 
We argued that this result can be interpreted as a signature of an inconsistency of general relativity in non-Euclidean topologies (see Sec.~4.3 in \citep{2022_Vigneron_c}). We then raised the following question: \textit{What relativistic equation admitting a non-relativistic limit in any topology should we consider?} The main requirements we drew for the new relativistic equation were the following: (i) It should reduce to the Einstein equation in a Euclidean topology; (ii) must be second order in the metric derivatives.
In that same paper \citep{2022_Vigneron_c}, we proposed an answer to the above question in the form of a bi-connection theory similar to the one introduced by Rosen~\citep{1980_Rosen}. It is composed of one physical Lorentzian structure~$(\T g, \T\nabla)$ and one non-dynamical reference connection $\T{\bar\nabla}$. The equations in this theory are the same as in~\citep{1980_Rosen}, in particular, the Einstein equation is modified such that the physical spacetime Ricci curvature $R_{\mu\nu}$ is replaced by the difference between that curvature and the reference Ricci curvature $\bar R_{\mu\nu}$ arising from the reference connection [see Eq.~\eqref{eq:biCoEq}]. The fundamental difference between Rosen's theory and the approach of \citep{2022_Vigneron_c} is in the choice of reference connection which \citep{2022_Vigneron_c} takes to be related to the spacetime topology. This theory only differs from general relativity in the case of non-Euclidean topologies, for which $\bar R_{\mu\nu} \not= 0$, and should be considered instead of the latter if one wants to study a model universe compatible with the non-relativistic regime in any topology.\saut

The goal of the present paper is to derive the equations of the cosmological model that result from this bi-connection theory (presented in Section~\ref{sec::bi-connection}) and confront them with observational data.  Within the Standard Model of Cosmology, three main sets of equations are used:
\begin{enumerate}
	\item[(i)] The homogeneous and isotropic solution of the Einstein equation to describe global expansion.
	\item[(ii)] The weak field limit to describe the linear regime of inhomogeneities in the early Universe, and in the late Universe on large scales. These equations allow us to test the model using the Cosmic Microwave Background (CMB) data, Baryonic Acoustic Oscillation (BAO) data, and Supernovae (SN1a) data in particular.
	\item[(iii)] The non-relativistic equations (cosmological Newton equations) to describe non-linear structure formation in the late Universe. $N$-body simulations performed using these equations allow us to test the model by comparing mock catalogs with catalogs of galaxies.
\end{enumerate}
These sets of equations need to be derived within the framework of the bi-connection theory for a complete cosmological model.\saut

The non-relativistic equations resulting from the bi-connection theory were already derived in \citep{2022_Vigneron_b, 2023_Vigneron_et_al}: for Euclidean and non-Euclidean topologies, they correspond to the cosmological Newton equations and the non-Euclidean Newtonian equations, respectively. The latter theory describes Newtonian (i.e. non-relativistic) gravitation on non-Euclidean topologies (e.g. spherical, hyperbolic, etc). It is shown in \citep{2023_Vigneron_et_al} how it could be used to study non-linear structure formation in spherical topologies.\saut

To complete the cosmological model related to the bi-connection theory of \citep{2022_Vigneron_c}, it remains to derive the homogeneous and isotropic solution (to describe expansion), along with the weak field limit of the bi-connection theory (to describe the linear regime of inhomogeneities). The former is derived in Sec.~\ref{sec::Exp_law}, where we show, in particular, that the curvature parameter $\Omega_K$ is not present anymore in the expansion law (i.e. ${\Omega_{\not= K} = 1, \ \forall \Omega_K}$) compared to the same solution derived from the Einstein equation (i.e. $\Omega_{\not= K} + \Omega_K = 1$): expansion is \textit{blind} to the spatial curvature. As a complementary result, we show in Appendix~\ref{eq::Exp_NEN} that this expansion law also holds for a general non-perturbative inhomogeneous solution in the non-relativistic limit. The weak field limit of the bi-connection theory is derived in Sec.~\ref{sec::Weak_field_limit} where we show that, as for the background solution, the presence of the curvature parameter in the equations is significantly changed compared to the Standard Model.\saut

On scales where non-linearities are important, the effects of the background spatial curvature and topology are expected to be smaller than current observational precision. Since the difference between general relativity and the bi-connection theory developed in~\citep{2022_Vigneron_c} is related to these two parameters, we expect observational differences to appear only on large scales, i.e. on scales described by the linear approximation. Therefore, while an $N$-body simulation seems not relevant to test the cosmological model related to the bi-connection theory\footnote{For this same reason a parameterised-post-Newtonian calculation, aimed at testing modified gravity theories on solar system scales, would not be relevant to test the bi-connection theory.}, a direct comparison with CMB data (in particular) using the weak field equations derived in Sec.~\ref{sec::Weak_field_limit} would provide a first test of this new theory. This test is performed in Sec.~\ref{sec:obs}. We conclude in Sec.~\ref{sec::disc}.

\section{The bi-connection theory of \citep{2022_Vigneron_c}}
\label{sec::bi-connection}

The bi-connection theory introduced in \citep{2022_Vigneron_c} is defined on a 4-manifold $\CM = \mathbb{R}\times\Sigma$ where $\Sigma$ is a closed 3-manifold, which we equip with
\begin{itemize}
	\item a physical Lorentzian metric $\T g$ and its connection $\T\nabla$. It defines the physical (spacetime) Riemann tensor $R^\mu{}_{\alpha\beta\nu}$, the physical Ricci tensor $R_{\mu\nu} \coloneqq R^\alpha{}_{\mu\alpha\nu}$ and the physical scalar curvature $R \coloneqq g^{\mu\nu} R_{\mu\nu}$.
	\item a non-dynamical reference connection $\bar{\T\nabla}$. It defines the reference (spacetime) Riemann tensor $\bar R^\mu{}_{\alpha\beta\nu}$ and the reference Ricci tensor $\bar R_{\mu\nu} \coloneqq \bar R^\alpha{}_{\mu\alpha\nu}$. No reference scalar curvature can be defined from $\T {\bar\nabla}$ alone.
\end{itemize}
The reference connection $\bar{\T \nabla}$ is non-dynamical in the sense that it is the same for any physical metric and energy-momentum tensor. In the approach of~\citep{2022_Vigneron_c}, that connection depends on topological properties of $\CM$ in the sense that it is chosen to be related to the universal cover $\tilde\CM = \mathbb{R}\times\tilde\Sigma$ of $\CM$, where $\tilde\Sigma$ is the universal cover of $\Sigma$. The universal cover does not determine the precise topology of $\CM$, but only its class. Since we always consider globally hyperbolic spacetimes (i.e. $\CM = \mathbb{R}\times\Sigma$), the choice of spacetime universal cover $\tilde\CM$ is equivalent to the choice of spatial universal cover $\tilde\Sigma$.\saut

The choice of $\bar{\T \nabla}$ made  in \citep{2022_Vigneron_c} is the following: we assume that there exists a coordinate system~$\{x^0,x^i\}$ adapted to a foliation of $\Sigma$-hypersurfaces such that the reference Riemann tensor writes\footnote{Throughout this paper, we denote indices running from 0 to 3 by Greek letters and indices running from 1 to 3 by Roman letters.}
\begin{equation}
	\bar R^\mu{}_{\alpha\nu\beta} = \delta^\mu_a \delta^i_\alpha \delta^b_\nu \delta^j_\beta \  {^{\tilde\Sigma}\bar{\CR}}^a{}_{ibj}(x^k), \label{eq::Riembar_choice}
\end{equation}
where ${^{\tilde\Sigma}\bar{\CR}}^a{}_{ibj}$ is independent of $x^0$ and corresponds to the standard Riemann tensor of the covering space $\tilde\Sigma$. In the cases of interest for the present paper, $\tilde\Sigma$ will either be the Euclidean $\mE^3$, the spherical $\mS^3$, or the hyperbolic $\mH^3$ covering spaces, but in general five other types of topologies are possible, as described by the Thurston decomposition~\citep{1995_La_Lu}. In these three cases, we, respectivel,y have
\begin{align}
    {^{\mE^3}\bar{\CR}}^a{}_{ibj} = 0 \quad ; \quad {^{\mS^3}\bar{\CR}}^a{}_{ibj} = \delta^a_b \bar h^{\mS^3}_{ij} - \delta^a_j \bar h^{\mS^3}_{ib} \quad ; \quad {^{\mH^3}\bar{\CR}}^a{}_{ibj} = \delta^a_b \bar h^{\mH^3}_{ij} - \delta^a_j \bar h^{\mH^3}_{ib},
\end{align}
where $\bar h^{\mS^3}_{ij}$ (respectively, $\bar h^{\mH^3}_{ij}$) are homogeneous and isotropic metrics on $\mS^3$ (respectively,~$\mH^3$). Therefore, in this paper, the reference Ricci tensor has the form
\begin{align}
	\bar R_{\mu\nu} = 2K \delta^i_\mu \delta^j_\nu \bar h_{ij}(x^k), \label{eq:current_choice}
\end{align}
with $\bar h_{ij}$ a homogeneous and isotropic metric, and $K=0$, $1$, and $-1$ for, respectively, Euclidean, spherical, and hyperbolic topologies. From this formula, ${\rm dim}\left[{\rm ker}\bar R_{\mu\nu}\right] = 3$ (in the spherical and hyperbolic cases), and therefore $\bar R_{\mu\nu}$ defines a reference observer of 4-velocity $G^\mu$ such that~$G^\mu \bar R_{\mu\nu} = 0$. The presence of this vector will be important for Sec.~\ref{sec::homo_iso}. Furthermore, as will be shown in that section, the normalisation factor of the reference Ricci curvature is a gauge choice.\saut

In this theory, the Einstein equation is modified to feature the reference curvature as follows: 
\begin{equation}
	G_{\alpha\beta} = \kappa\, T_{\alpha\beta} - \Lambda g_{\alpha\beta} + \Top_{\alpha\beta}, \label{eq::ModE1}
\end{equation}
where $\kappa \coloneqq 8\pi G$, $T_{\alpha\beta}$ is the energy-momentum tensor, $\Lambda$ the cosmological constant, and $\Top_{\alpha\beta}$ is defined as
\begin{equation}
	\Top_{\alpha\beta} \coloneqq \bar{R}_{\alpha\beta} - \frac{\bar{R}_{\mu\nu}g^{\mu\nu}}{2} g_{\alpha\beta}. \label{eq::Top_munu}
\end{equation}
Since the reference curvature directly depends on the spacetime topology by the choice~\eqref{eq::Riembar_choice}, the term $\Top_{\alpha\beta}$ can be considered to be a \textit{topological term}.
Equation~\eqref{eq::ModE1} can be rewritten in the more convenient form
\begin{align}
	R_{\alpha\beta} - \bar{R}_{\alpha\beta} = \kappa \, \left(T_{\alpha\beta} - \frac{T^\mu{}_\mu}{2} g_{\alpha\beta} \right) + \Lambda g_{\alpha\beta}. \label{eq:biCoEq}
\end{align}
We see from this equation that the difference with general relativity is to replace the physical spacetime Ricci tensor with the difference between that tensor and the reference spacetime Ricci tensor. The main interpretation of that equation is that matter does not curve spacetime anymore, as in general relativity, but only induces a departure of the physical Ricci curvature from the reference, topological, Ricci curvature.\saut

The additional term $\Top_{\alpha\beta}$ in the Einstein equation is conserved:
\begin{align}
	g^{\mu\nu}\left(\nabla_\mu \bar R_{\nu\alpha} - \frac{1}{2} \nabla_\alpha \bar R_{\mu\nu}\right) = 0. \label{eq:biCoCond}
\end{align}
This equation, called the \textit{bi-connection condition}, constrains the diffeomorphism freedom in the definition of $\bar R_{\mu\nu}$ with respect to $g_{\mu\nu}$.\saut

Equations~\eqref{eq:biCoEq} and \eqref{eq:biCoCond} are equivalent to the ones of the bi-connection theory proposed  by \citet{1980_Rosen}. The only, but fundamental, difference is the choice and motivation for the reference connection: Rosen chose a reference connection related to a de Sitter metric in order to remove singularities from general relativity, while in our case, the reference connection is topology dependent as it is related to the universal cover of the spacetime {manifold}~$\CM$. \saut

In the case of a Euclidean topology, i.e. $\tilde\Sigma = \mE^3$, we have $\bar R_{\mu\nu} = 0$, implying that Eq.~\eqref{eq:biCoEq} is equivalent to the Einstein equation and that Eq.~\eqref{eq:biCoCond} is trivial. Therefore, general relativity and the bi-connection theory of \citep{2022_Vigneron_c} coincide for Euclidean topologies, and differ for any other type of topology. In terms of the cosmological model, this will imply that the two theories will differ only if~$\Omega_K \not= 0$.

\section{Homogeneous and isotropic solution of the bi-connection theory}
\label{sec::Exp_law}

\subsection{Derivation}
\label{sec::homo_iso}

We assume $\T g$ to be the Friedmann-Lema\^itre-Robertson-Walker (FLRW) metric. Therefore, the covering space $\tilde\Sigma$ is either $\mE^3$, $\mS^3$ or $\mH^3$, and the reference Ricci curvature has the form~\eqref{eq:current_choice}. We define $\T n$ to be the 4-velocity of the observer relative to the homogeneous foliation of $\T g$, for which the spatial metric is denoted $\T h$. Note that at this stage, $\bar h_{ij}$ present in~\eqref{eq:current_choice} is {\it a priori} different from $h_{ij}$, and these two tensors are not necessarily related to the same foliation. The relation between them will be constrained by Eq.~\eqref{eq::ModE1}.\saut

Because of homogeneity and isotropy, we can write both $T_{\mu\nu}$ and $\Top_{\mu\nu}$ as\footnote{The most general solution {\it a priori} features heat fluxes $q_\alpha$ and $\bar q_\alpha$ from, respectively, $T_{\alpha\beta}$ and $\Top_{\alpha\beta}$ constrained to be $\bar q_\alpha = -q_\alpha$ by Eq.~\eqref{eq::ModE1}. This corresponds to a tilted cosmological model: both the fluid and the reference observer defined by $G^\alpha$ are tilted with respect to the homogeneous foliation.}
\begin{align}
	T_{\alpha\beta} = \rho n_\alpha n_\beta + p h_{\alpha\beta}, \\
	\Top_{\alpha\beta} = \bar\rho n_\alpha n_\beta + \bar p h_{\alpha\beta}, \label{eq::def_bar_rho_p}
\end{align}
where $\rho$ and $p$ are, respectively, the energy density and pressure of matter, and
\begin{align}
	\bar\rho &\coloneqq \frac{1}{2} \left(n^\mu n^\nu \bar R_{\mu\nu} + h^{\mu\nu} \bar R_{\mu\nu}\right),\\
	3\bar p &\coloneqq \frac{1}{2} \left(3n^\mu n^\nu \bar R_{\mu\nu}- h^{\mu\nu} \bar R_{\mu\nu}\right) \label{eq::bar_p}
\end{align}
are the effective energy density and pressure coming from the topological term. Then the expansion laws take the form
\begin{align}
	3H^2		&= \kappa\rho + \bar\rho + \Lambda - \CR/2, \label{eq::H21} \\
	3\ddot a/a	&= -\frac{\kappa}{2}\left(\rho + 3 p\right) - \frac{1}{2}\left(\bar\rho + 3\bar p\right) + \Lambda, \label{eq::ddota1}
\end{align}
where $\CR = 6K/a^2$ is the scalar spatial curvature related to the physical spatial metric $\T h$, with $a(t)$ the scale factor and $H = \dot a / a$ the expansion rate. It remains to find a more explicit formula for $\bar\rho$ and $\bar p$.\saut

The heat flux relative to the term $\T\Top$ being zero implies $n^\mu h^{\alpha\nu} \bar R_{\mu\nu} = 0$. Coupled with the fact that ${\rm dim}\left[{\rm ker} \bar R_{\mu\nu}\right] = 3$ from relation~\eqref{eq:current_choice}, then $n^\mu \bar R_{\mu\nu} = 0$. This implies that the observer related to the homogeneity foliation induced by the FLRW metric corresponds to the reference observer induced by the reference spacetime curvature, i.e. $G^\mu \propto n^\mu$. Then, using~\eqref{eq::def_bar_rho_p}--\eqref{eq::bar_p} along with $\Top_{\alpha\beta} \coloneqq \bar R_{\alpha\beta} - \frac{\bar R_{\mu\nu}g^{\mu\nu}}{2}g_{\alpha\beta}$, we get 
\begin{align}
	\bar \rho &= -3\bar p = \frac{1}{2}h^{\mu\nu}\bar R_{\mu\nu}, \\
	\bar R_{\alpha\beta} &= -2\bar p \, h_{\alpha\beta}.
\end{align}
%Using the bi-connection condition~\eqref{eq::metrics_cond1}, corresponding to the conservation of $\T \Top$, we have
%\begin{align}
%	\dot{\bar{\rho}}	&= -3H\left(\bar\rho + \bar p\right) = 2H\bar\rho.
%\end{align}
%Then $\bar\rho = -3\bar p \propto 1/a(t)^2$.\saut
In coordinates adapted to the foliation of homogeneity, the second relation, along with~\eqref{eq:current_choice}, leads to
\begin{equation}
	2K\bar h_{ij} = -2\bar p h_{ij}. \label{eq::barh_ij}
\end{equation}
Both $K$ and $\bar p$ being spatial constants, the above equation implies
\begin{align}
    \bar\CR_{ij} = \CR_{ij}, \label{dd}
\end{align}
where $\bar\CR_{ij} = 2K\bar h_{ij}$ is the Ricci tensor associated with $\bar h_{ij}$. Furthermore, for $K\not=0$, the inverse of~\eqref{eq::barh_ij} leads to
\begin{equation}
	\bar h^{ij} = -\frac{K}{\bar p} h^{ij}, \label{eq::barh^ij}
\end{equation}
where $\bar h^{ij}$ is the inverse of $\bar h_{ij}$ (i.e.  $\bar h^{ij} \not= \bar h_{cd} h^{ci} h^{dj}$).  Then, using Eq.~\eqref{dd} we get $\CR \coloneqq \CR_{ij} h^{ij} = \bar \CR_{ij} h^{ij}$, which, along with relation~\eqref{eq::barh^ij}, leads to
\begin{equation}
	6\bar p = -\CR.
\end{equation}
This implies $\bar\rho = \CR/2$. Finally, the expansion laws~\eqref{eq::H21} and~\eqref{eq::ddota1} of an exact homogeneous and isotropic solution of the bi-connection theory are
\begin{align}
	3H^2		&= \kappa\rho + \Lambda, \quad \forall K \label{eq::H2_bi-connection} \\
	3\ddot a/a	&= -\frac{\kappa}{2}\left(\rho + 3 p\right) + \Lambda. \label{eq::ddota_bi-connection}
\end{align}
These expansion laws are the ones of a flat homogeneous and isotropic model as derived with the Einstein equation, but here they hold even in the non-flat cases, i.e. for all $K$.\saut

While the bi-connection theory has the additional field $\bar R_{\mu\nu}$ with respect to general relativity, we see that the exact homogeneous and isotropic solution does not have an additional parameter linked to this field. The only role of $\bar R_{\alpha\beta}$ is to set the topology, the equations being independent of the value chosen for the reference scalar curvature $\bar\CR \coloneqq \bar h^{ij} \bar \CR_{ij}$. Indeed, rescaling the choice~\eqref{eq:current_choice} by a constant factor, which would rescale $\bar\CR$ by the same factor, only results in a rescaling of the scale factor $a(t)$. Therefore, the value of $\bar\CR$ is just a gauge choice. This was not the case with Rosen's choice of reference curvature \citep{1980_Rosen}, where a reference cosmological constant was introduced.

\subsection{Why is it expected?}

Equation~\eqref{eq::H2_bi-connection} shows that within the framework of the present bi-connection theory, the expansion scenario is the same for a Euclidean, spherical, or hyperbolic universe. While we discuss the consequences of this result in more details in Sec.~\ref{sec:blind}, in the present section we explain why it is expected from any relativistic theory which we require to have a non-relativistic limit in any topology.\saut

Let us consider the first Friedmann equation resulting from the Einstein equation, where we reintroduce the speed of light $c\not=1$:
\begin{equation}
		3H^2 - \kappa\rho - \Lambda + c^2{\CR}/{2} = 0.
\end{equation}
We see that the curvature term appears as a $-1$ order in $1/c^2$, while the other terms are all zeroth order terms. Requiring the non-relativistic limit to exist corresponds to requiring that this equation be written as a Taylor series of $1/c^2$, and therefore that each order needs to be independently zero. This implies
\begin{align}
	\frac{\CR(t)}{2}	&= 0, \quad\quad\quad\; \, \textrm{(order $-$1)} \label{eq::bite2} \\
	3H^2			&= \kappa\rho + \Lambda. \quad \textrm{(order 0)} \label{eq::bite1}
\end{align}
Therefore, the solution necessarily needs to describe a flat universe. This is a rough derivation of the result in \citep{2022_Vigneron_c} for the specific case of a homogeneous and isotropic solution, stating that no non-relativistic limit of the Einstein equation exists for a solution describing a non-Euclidean spatial topology.\saut

The role of the reference spacetime curvature added in the Einstein equation is to allow for this limit to be possible. In the present case of a homogeneous solution, the term~$\T\Top$ adds an effective density in the Friedmann equations, which, as shown in Sec.~\ref{sec::homo_iso}, cancels the spatial curvature term. The consequence is that the expansion law does not feature the negative order in $1/c^2$ anymore, and therefore, from the Taylor series, we only obtain~\eqref{eq::bite1} without the zero curvature constraint~\eqref{eq::bite2}, i.e. without constraining the topology to be Euclidean. For this reason, we expect the expansion law~\eqref{eq::H2_bi-connection} to hold for any relativistic theory admitting a  non-relativistic limit in any topology, i.e. not only with the bi-connection theory of \citep{2022_Vigneron_c}.

\subsection{Expansion is blind to the spatial curvature}
\label{sec:blind}

The expansion laws in our cosmological model [Eqs.~\eqref{eq::H2_bi-connection} and \eqref{eq::ddota_bi-connection}] are the one of a flat Lambda-Cold-Dark-Matter ($\Lambda$CDM) model, \textit{regardless of the spatial curvature}: 
\begin{equation}
	\Omega  = 1, \quad \forall \, \Omega_K \label{eq::Omega_law}
\end{equation}
where $\Omega \coloneqq \Omega_{\rm m} + \Omega_{\rm r}  + \Omega_\Lambda$, with $\Omega_{\rm m} \coloneqq \kappa\rho_{\rm m}/(3H^2)$ the matter parameter, ${\Omega_{\rm r} \coloneqq \kappa\rho_{\rm r}/(3H^2)}$ the radiation parameter, $\Omega_K \coloneqq -K/(a^2H^2)$ the curvature parameter and ${\Omega_\Lambda \coloneqq \Lambda/(3H^2)}$ the cosmological constant parameter.  This result also holds in the presence of inhomogeneities and non-linearities if these are non-relativistic (see Appendix~\ref{eq::Exp_NEN}).
Therefore, the expansion as predicted by the bi-connection theory is {\it blind} to the spatial curvature, while this curvature still affects the measure of distances. Therefore, in the bi-connection theory, spatial curvature has smaller effects on the dynamics than in general relativity; the effect remains essentially geometrical. This is a strong difference between the two theories, which leads to two main questions:
\begin{itemize}
	\item What is the value of the curvature parameter resulting from a reanalysis of the cosmological data with relation~\eqref{eq::Omega_law}? 
	\item In the case where the reevaluated curvature parameter is not negligible anymore, are the values of other cosmological parameters changed such that recent observational tensions within the $\Lambda$CDM model can be solved?
\end{itemize}

The first question is especially interesting in light of a rising debate on the value of the spatial curvature that should be inferred from the CMB data of the Planck space observatory. As shown in \citep[e.g.][]{2020_Di-Valentino_et_al, 2020_Efstathiou_et_al, 2021_Handley, 2021_Di-Valentino_et_al_bis, 2021_Vagnozzi_et_al_a, 2021_Vagnozzi_et_al_b, 2021_Dhawan_et_al}, the best fit of the Planck CMB power spectrum at all scales seems to prefer\footnote{Although, this depends on the likelihood used. With the Planck ``Camspec'' (and the new NPIPE) \citep{2020_Efstathiou_et_al,2023_Tristram_et_al}, there is less preference for a curved universe than the conventional ``plik'' likelihood \citep{2020_Planck_VI}.} a value at current time of $\Omega_{K,0} \simeq -0.045$% (equivalently in their case $\Omega_{K,0} h^2 \simeq -0.013$, with $h = H_0/(100 \ \rm{km/s/Mpc})$)
, which differs from the standard constraint $|\Omega_{K,0}| \lesssim 10^{-3}$ obtained when BAO data are taken into account\footnote{Let us mention that because of the tension between Planck and BAO data regarding curvature, it has been argued that it may not be statistically consistent to combine both datasets \cite{2021_Handley}.} \citep{2020_Planck_VI}. The main issue with this result is that it leads to a value of the Hubble constant $H_0^{\rm CMB}$ inferred from the CMB that is particularly low, strongly increasing the tension with the local supernovae measurement $H_0^{\rm SN1a}$ \citep{2022_Riess_et_al}: from $H_0^{\rm SN1a} - H_0^{\rm CMB} \sim 5$ km/s/Mpc to $H_0^{\rm SN1a} - H_0^{\rm CMB} \sim 20$ km/s/Mpc (with large error bars). Since expansion is not directly affected by spatial curvature in our model, one may expect that the preference for non-zero $\Omega_K$ when fitting the CMB alone could stay without changing the Hubble constant, and thus not increasing the Hubble tension.\saut

The reason why $H_0$ must change when $\Omega_K \not=0$ in an analysis of CMB data comes mainly from the angular diameter distance 
\begin{align}
	d_{\rm A}(z) = \frac{1}{\sqrt{|\Omega_{K,0}|}H_0 (1+z)} S_K\left(\sqrt{|\Omega_{K,0}|}H_0\int_0^z\frac{\mathrm{d} z'}{H(z')}\right), \label{eq:dA}
\end{align}
where
\begin{align}
	S_K(x) =
		\begin{cases}
			&\sin x,	\quad \Omega_K < 0 \\
			&x ,		\quad \Omega_K = 0 \\
			&\sinh x,	\quad \Omega_K > 0.
		\end{cases}
\end{align}
When evaluated at the recombination redshift, the angular diameter distance sets the typical distance under which we observe CMB angular anisotropies. The typical angular scale of CMB anisotropies (the ``sound horizon'' $\theta_s$) is very well determined by Planck to subpercent precision \cite{2020_Planck_VI}. As  $\Omega_K$ scales like $(1+z)^2$, it does not affect the early Universe, but can affect the angular diameter distance, hereby changing $\theta_s$. 
The degeneracy with $H_0$ allows us to compensate for the effect of $\Omega_K$ at the background level.\saut

In fact, this formula is common to both the $\Lambda$CDM model and our model, but with $H(z)$ solution of either ${\Omega + \Omega_K = 1}$ ($\Lambda$CDM model), or ${\Omega = 1\ \ \forall \Omega_K}$ (our model). Therefore, for the Standard Model, spatial curvature appears in two places in formula~\eqref{eq:dA}: as a geometrical effect in the function $S_K(x)$, and as a dynamical effect in $H(z)$; while in our model, it only appears as a geometrical effect in $S_K(x)$. Consequently, if the shift from $H_0^{\rm SN1a} - H_0^{\rm CMB} \sim 5$ km/s/Mpc to $H_0^{\rm SN1a} - H_0^{\rm CMB} \sim 20$ km/s/Mpc obtained in \citep[e.g.][]{2020_Di-Valentino_et_al, 2021_Handley, 2021_Di-Valentino_et_al_bis} comes mainly from the presence of $\Omega_K$ in $S_K(x)$, this shift should still be present in our model. However, if it comes from the presence of $\Omega_K$ in $H(z)$, this shift should disappear in our model, and the curvature tension might be solved.\saut

To properly determine which case we are in, and to answer the above two questions, a full analysis of the CMB data using a Boltzmann code is necessary. It is performed in Sec.~\ref{sec:obs}. This requires the derivation of the first order perturbation equations of the bi-connection theory, which are presented in the next section.

\section{Weak field limit}
\label{sec::Weak_field_limit}

\subsection{Gauge invariant variables and equations}

In this section, we recall the definitions and the gauge invariant equations used in the weak field limit. We follow the notation of \citep{2009_Malik_et_al}. The limit is a first order perturbation of a FLRW metric $g_{\mu\nu} = g^{\rm FLRW}_{\mu\nu} + \delta g_{\mu\nu}$ with
\begin{align}
	{g}^{\rm FLRW}_{\mu\nu} = a^2\left(\begin{array}{cc}
		-1 & 0\\
		0 &   h_{ij}\end{array}\right), \label{eq::phys_g}
\end{align}
and
\begin{align}
	\delta{g}_{\mu\nu} = a^2\left(\begin{array}{cc}
		-2\phi & D_i B -S_i\\
		D_i B -S_i &  -2\psi  h_{ij} + 2 D_{i} D_jE + 2 D_{(i}F_{j)} + 2f_{ij}\end{array}\right), \label{eq::eta_munu}
\end{align}
where $D_c S^c \coloneqq 0$, $D_ cF^c \coloneqq 0$, $f_c{}^c \coloneqq 0$, and $D_c f^{ci} \coloneqq 0$. The FLRW metric is written in conformal time $\tau$. The expansion rate is denoted $\CH = a'/a$ where the prime derivative is with respect to conformal time. We stress that in the present convention (which we consider for all of Sec.~\ref{sec::Weak_field_limit}), $h_{ij}$ is the comoving spatial metric with $D_ i$ its connection, and all the spatial indices are raised and lowered with that metric.\saut

Under a gauge transformation, which can be described by an infinitesimal change of coordinates $x^\mu \rightarrow x^\mu + \xi^\mu$, the gauge invariant variables related to the metric are
\begin{align}
	\Psi &\coloneqq \psi + \CH\sigma, \\
	\Phi &\coloneqq \phi - \CH\sigma - \sigma', \\
	Q_i &\coloneqq S_i + F_i',
\end{align}
with $\sigma \coloneqq E' - B$. The tensor mode $f_{ij}$ is already gauge invariant.\saut

The components of the energy-momentum tensor take the form
\begin{align}
	T^0{}_0	&= -(\rho + \delta\rho), \label{eq::T^0_0}\\
	T^0{}_i	&= (\rho + p)(D_i v + v_i + D_i B - S_i), \\
	T^i{}_j	&= (p+\delta p)\delta^i_j + (D^iD_j - \frac{1}{3}\delta^i_j\Delta)\Pi + \frac{1}{2}(D^i \Pi_j + D_j\Pi^i) + \Pi^i{}_j, \label{eq::T^i_j}
\end{align}
where $\rho$ and $p$ are the homogeneous energy density and pressure, $\Pi$, $\Pi_i$, and $\Pi_{ij}$ are, respectively, the scalar, vector, and tensor parts of the anisotropic stress. The gauge invariant quantities related to the energy-momentum tensor are
\begin{align}
	\rhoNew &\coloneqq \delta\rho -\rho'\sigma, \label{eq::delta_rho_sigma}\\
	V &\coloneqq v + E', \\
	\delta p_{\not= \rm ad} &\coloneqq \delta p - c_{\rm s}^2\delta\rho, \\
	%V_i &\coloneqq v_i + F_i', \\
	q_i &\coloneqq (\rho + p)(v_i - S_i), \label{eq::q}
\end{align}
with $c_{\rm s}^2 \coloneqq p'/\rho'$. The anisotropic stress variables are already gauge invariant. We also define $\delta \coloneqq \rhoNew/\rho$. 

Then, the first order gauge invariant equations from the Einstein equation $G_{\mu\nu} = \kappa T_{\mu\nu}$, in which we include $\Lambda$ in $T_{\mu\nu}$, are

\begin{enumerate}
\item for scalar modes:
\begin{align}
	&(\Delta + 3K)\Psi = a^2\frac{\kappa}{2}\rho \delta + 3\CH\left(\Psi' + \CH\Phi\right), \label{eq::EE_S_mode_1}\\
	&\Psi' + \CH \Phi = -a^2\frac{\kappa}{2}(\rho+p)V, \\
	&\Psi'' + 2\CH\Psi' + \CH\Phi' + (2\CH' + \CH^2)\Phi - K\Psi = a^2\frac{\kappa}{2}\left(c_{\rm s}^2\rho \delta + \delta p_{\not= \rm ad} + \frac{2}{3}\Delta\Pi\right), \\
%&\Psi'' + \CH(2 + 3c_{\rm s}^2)\Psi' + \CH\Phi' + \left[2\CH' + (1+3c_{\rm s}^2)\CH^2\right]\Phi\ - K\Psi \nonumber\\
	%&\qquad= a^2\frac{\kappa}{2}\left[c_{\rm s}^2\rho D_\rho + \delta p_{\not= \rm ad} + \frac{2}{3}\Delta\Pi\right], \\
	&\Psi - \Phi = a^2\kappa\Pi, \label{eq:gshift}
\end{align}

\item for vector modes:
\begin{align}
	\left(\Delta + 2K\right)Q_i &= -a^2 2\kappa (\rho + p)(V_i - Q_i), \\
	Q_i' + 2\CH Q_i &= a^2\kappa\Pi_i,
\end{align}

\item for tensor modes:
\begin{align}
	f_{ij}'' + 2\CH f_{ij}' + \left(2K - \Delta\right)f_{ij} = a^2\kappa \Pi_{ij}. \label{eq::EE_T_mode}
\end{align}
\end{enumerate}
The first order gauge invariant equations from the conservation law $\nabla_\nu T^{\mu\nu} = 0$ are:

\begin{enumerate}
\item for scalar modes:
\begin{align}
    &\delta' + 3\CH\left[\left(c_{\rm s}^2 - w\right)\delta + \delta p_{\not= \rm ad}/\rho\right] + (1+w)\left(\Delta V - 3\Psi'\right) = 0, \label{eq::cons_1}\\
	%&\rhoNew' + 3\CH\left[(1 + c_{\rm s}^2)\rhoNew + \delta p_{\not= \rm ad}\right] + (\rho + p)\Delta V - 3(\rho + p)\Psi' = 0, \label{eq::cons_1}\\
	%&D_\rho' - 3HwD_\rho = (\Delta + 3K)\left[-(1+w)V + 2\CH \Pi/\rho\right], \label{eq::cons_1}\\
	&V' + \CH\left(1-3c_{\rm s}^2\right) V + \Phi + \frac{1}{\rho + p}\left[c_{\rm s}^2\rho \delta + \delta p_{\not= \rm ad} + \frac{2}{3}\left(\Delta + 3K\right)\Pi\right] = 0, \label{eq::cons_2}
	%&V' + \CH V + \Phi + \frac{1}{\rho + p}\left[c_{\rm s}^2\rho D_\rho + \delta p_{\not= \rm ad} + \frac{2}{3}\left(\Delta + 3K\right)\Pi\right] = 0, \label{eq::cons_2}
\end{align}

\item for vector modes:
\begin{align}
	q_i' + 4\CH q_i =  -\frac{1}{2}\left(\Delta + 2K\right)\Pi_i, \label{eq::cons_V_mode}
	%q_i' + 4\CH q_i =  -\left(\Delta + 2K\right)\Pi_i, \label{eq::cons_V_mode}
\end{align}
where we defined $w \coloneqq p/\rho$. In a non-interacting multi fluid approach, these last three equations are fulfilled for each fluid component.
\end{enumerate}
%We also give the equation for the growth of fluid rest frames density perturbations
%\begin{align}
%	&D_\rho'' + \CH\left(1+3c_{\rm s}^2 - 6w\right)D_\rho' + \left[\left(\CH^2 + K\right)\left(\frac{9}{2}w^2 - 12w + 9c_{\rm s}^2 - \frac{3}{2}\right) \right.  \label{eq::growth_EE} \\
%	&\quad  + \left(12w - 9c_{\rm s}^2\right)K - c_{\rm s}^2\left(\Delta+3K\right) \Big]D_\rho = w\left(\Delta+3K\right) \frac{\delta p_{\not= \rm ad}}{p} + 2\CH w\left(\Delta + 3K\right)\left(\frac{\Pi}{p}\right)' \nonumber \\
%	&\quad + 2\left[\left(\CH^2 + K\right)\left(3c_{\rm s}^2 - 2w + 3w^2\right) + \left(4w - 3c_{\rm s}^2\right)K - \frac{w}{3}\left(\Delta+3K\right) \right]\left(\Delta + 3K\right)\frac{\Pi}{p}.\nonumber
%\end{align}

The goal of the next sections is to derive the same gauge invariant equations as~\eqref{eq::EE_S_mode_1}--\eqref{eq::cons_V_mode} in the case where the Einstein equation features the topological term ${\Top_{\mu\nu} \coloneqq \bar R_{\alpha\beta} - \frac{\bar R_{\mu\nu}g^{\mu\nu}}{2}g_{\alpha\beta}}$. For this, we can treat $\Top_{\mu\nu}$ as an additional effective energy-momentum tensor with zeroth order quantities $\bar\rho$ and $\bar p$, and first order gauge invariant quantities $\delta\bar\rho_\sigma$, $\bar V$, $\delta\bar p_{\not= \rm ad}$, $\bar q_i$, $\bar\Pi$, $\bar\Pi_i$ and $\bar\Pi_{ij}$. Then, Eqs.~\eqref{eq::EE_S_mode_1}--\eqref{eq::EE_T_mode} are changed as follows:
\begin{align}
	&\rhoNew \rightarrow \rhoNew + \frac{1}{\kappa}\delta\bar\rho_\sigma \quad;\quad(\rho + p)V \rightarrow (\rho + p)V + \frac{1}{\kappa}(\bar\rho + \bar p)\bar V \quad; \nonumber \\
	&\delta p_{\not= \rm ad} \rightarrow \delta\bar p_{\not= \rm ad} + \frac{1}{\kappa}\delta\bar p_{\not= \rm ad} \quad;\quad q_i \rightarrow  q_i + \frac{1}{\kappa}\bar q_i \quad ; \label{eq::transfo} \\
	&\Pi \rightarrow \Pi + \frac{1}{\kappa}\bar \Pi \quad;\quad \Pi_i \rightarrow  \Pi_i + \frac{1}{\kappa}\bar \Pi_i \quad;\quad \Pi_{ij} \rightarrow  \Pi_{ij} + \frac{1}{\kappa}\bar \Pi_{ij}. \nonumber
\end{align}
It remains to find the first order quantities associated with the topological term.

\subsection{Gauge invariant quantities of the topological term}

As shown in Sec.~\ref{sec::Exp_law}, for a homogeneous and isotropic solution, in a coordinate system where the physical spacetime metric can be written as \eqref{eq::phys_g}, the reference spacetime curvature takes the form ${\bar R}_{\mu\nu}  = 2K\delta_\mu^i\delta_\nu^j h_{ij}$ (i.e. the reference spatial metric $\bar h_{ij}$ corresponds to the comoving spatial metric). In the framework of the weak field limit, this formula corresponds to the zeroth order of ${\bar R}_{\mu\nu}$. As presented in Sec.~\ref{sec::bi-connection}, that tensor is non-dynamical, i.e. it is fixed for a given topology and is not affected by the physics behind $T_{\mu\nu}$ and $g_{\mu\nu}$. Nevertheless, this does not mean that ${\bar R}_{\mu\nu}$ is only a zeroth order term within the weak field limit. Indeed, the first order of the physical metric~\eqref{eq::eta_munu} and the first order of the energy-momentum tensor \eqref{eq::T^0_0}--\eqref{eq::T^i_j} not only come from physics, but also from gauge freedom. This implies that ${\bar R}_{\mu\nu}$ has, in general, a non-zero first order term solely coming from gauge freedom, i.e.
\begin{align}
	{\bar R}_{\mu\nu} = \overset{0}{\bar R}_{\mu\nu} + \Lie{\T X}\overset{0}{\bar R}_{\mu\nu}, \label{eq::gauge_Rbar}
\end{align}
where $X^\mu$ is a first order 4-vector, $\Lie{\T X}$ is the Lie derivative along $\T X$, and $\overset{0}{\bar R}_{\mu\nu}  = 2K\delta_\mu^i\delta_\nu^j h_{ij}$. We direct the reader to Appendix~F of \citep{2022_Vigneron_c} for a more detailed justification of~\eqref{eq::gauge_Rbar}. This is done within the framework of the non-relativistic limit, but the derivation is equivalent to that with the weak field limit.\saut

%Another way of understanding the origin of the above formula is the following. From the knowledge of the physical spacetime metric $g_{\mu\nu}$, the decomposition into a background quantity $g^{\rm FLRW}$ and a perturbed quantity $\delta g_{\mu\nu}$ is unique up to (at first order) a Lie derivative $\Lie{\T\xi} g_{\mu\nu}$ (see \citep{2008_Malik_et_al}). From Section~\ref{sec::Exp_law}, we know that there exists one FLRW metric $\tilde g^{\rm FLRW}$ such that the reference spacetime curvature takes the form $\overset{0}{\bar R}_{\mu\nu}  = 2K\delta_\mu^i\delta_\nu^j \tilde h_{ij}$, with $\tilde h_{ij}$ the comoving spatial metric related to $\tilde g^{\rm FLRW}$. 

Because $\overset{0}{\bar R}_{\mu\nu}$ is purely spatial and its time derivative is zero, only the spatial components of $X^\mu$ remain in the first order. Therefore, we have
\begin{align}
	{\bar R}_{\mu\nu} = 2K\left(\begin{array}{cc}
		0 & D_i \CX' + \CX_i' \\
		D_i \CX' + \CX_i' &  h_{ij} + 2D_{i}D_j\CX + 2D_{(i}\CX_{j)}\end{array}\right),
\end{align}
where $ \delta_\mu^i X^\mu \eqqcolon D^i \CX + \CX^i$ with $D_c \CX^c \coloneqq 0$. Under a gauge transformation, we have
\begin{align}
	\bar R_{\mu\nu}		&\ \accentset{{\T \xi}}{\longrightarrow}\  \bar R_{\mu\nu} + \Lie{\T \xi}\bar R_{\mu\nu} \nonumber \\
	 				&\ \accentset{{\T \xi}}{\longrightarrow}\  \overset{0}{\bar R}_{\mu\nu} + \Lie{\T X + \T \xi}\overset{0}{\bar R}_{\mu\nu} \nonumber \\
	 				&\ \accentset{{\T \xi}}{\longrightarrow}\ 2K\left(\begin{array}{cc}
		0 & D_i (\CX' + \xi') + \CX_i' + \xi_i'\\
		D_i (\CX' + \xi') + \CX_i' + \xi_i' &  h_{ij} + 2D_{i}D_j(\CX + \xi) + 2D_{(i}\left[\CX_{j)} + \xi_{j)}\right]\end{array}\right),
\end{align}
where $\xi^\mu \delta_\mu^i \eqqcolon D^i \xi + \xi^i$ with $D_i \xi^i \coloneqq 0$. So, $\CX$ and $\CX^i$ are not gauge invariant but transform as, respectively, $\CX\rightarrow \CX + \xi$ and $\CX^i\rightarrow \CX^i+\xi^i$. Therefore, the following quantities defined from the reference Ricci curvature are gauge invariant:
\begin{align}
	\CC &\coloneqq \CX - E, \\
	\CC^i &\coloneqq \CX^i - F^i.
\end{align}
These variables are interpreted in Sec.~\ref{sec::interp_C}. We have
\begin{align}
	a^2\Top^0{}_0		&= -3K - 2K\Delta \CC - 6K\psi, \\
	a^2\Top^0{}_i		&= -2K(D_i \CC' +\CC_i' + D_i\sigma + Q_i), \\
	a^2\Top^i{}_j		&= -K\delta^i_j - 2K\delta^i_j \psi  + 4KD_i D^j \CC + 2K(D^i \CC_j + D_j\CC^i) - 4Kf^i_j.
\end{align}
Using the above equations along with \eqref{eq::T^0_0}--\eqref{eq::q}, the first order gauge invariant quantities (which we denote $\delta\bar\rho_\sigma$, $\bar V$, $\delta\bar p_{\not= \rm ad}$, $\bar q_i$, $\bar\Pi$, $\bar\Pi_i$, $\bar\Pi_{ij}$) defined from $\Top_{\mu\nu}$ are
\begin{align}
	&\delta\bar\rho_\sigma = \frac{2K}{a^2}(\Delta\CC + 3\Psi) \quad;\quad \bar V = -\CC' \quad;\quad \delta\bar p_{\not= \rm ad}	= 0 \quad;\quad \bar \Pi	= \frac{4K}{a^2}\CC \quad; \label{eq::GI_Top_S} \\
	&\bar q_i				= - \frac{2K}{a^2}\left(\CC'_i + Q_i\right)\quad;\quad \bar \Pi_i = \frac{4K}{a^2}\CC_i \quad;  \\
	&\bar \Pi_{ij}			= \frac{4K}{a^2}f_{ij}. \label{eq::GI_Top_T}
\end{align}
%and $\delta\bar\rho_\Delta = \frac{2K}{a^2}(\Delta\CC + 3\Psi + 3\CH \CC')$. 
We recall that the zeroth order quantities are derived in Sec.~\ref{sec::homo_iso} and are $\bar\rho = 3K/a^2$ and $\bar p = -K/a^2$.

\subsection{Gauge invariant equations of the bi-connection theory}
\label{sec::gauge_eq_bi_metric}

Introducing the gauge invariant quantities~\eqref{eq::GI_Top_S}--\eqref{eq::GI_Top_T} of the topological term in Eqs.~\eqref{eq::EE_S_mode_1}--\eqref{eq::cons_V_mode} as presented in~\eqref{eq::transfo}, the first order gauge invariant equations of the bi-connection equation~\eqref{eq::ModE1}, in which we include $\Lambda$ in $T_{\mu\nu}$, are:
\begin{enumerate}
\item for scalar modes:
\begin{align}
	%&\Delta\Psi = a^2\frac{\kappa}{2}\rho D_\rho + K\Delta\CC +  3K\CH\CC', \label{eq::Poisson_CC} \\
     &\Delta\Psi = a^2\frac{\kappa}{2}\rho \delta  + 3\CH\left(\Psi' + \CH\Phi\right) + K\Delta\CC , \label{eq::Poisson_CC} \\
	&\Psi' + \CH \Phi = -a^2\frac{\kappa}{2}(\rho+p)V + K\CC', \\
	&\Psi'' + 2\CH\Psi' + \CH\Phi' + (2\CH' + \CH^2)\Phi = a^2\frac{\kappa}{2}\left(c_{\rm s}^2\rho\delta + \delta p_{\not= \rm ad} + \frac{2}{3}\Delta\Pi\right) + K\Delta\CC, \\
    %&\Psi'' + \CH(2 + 3c_{\rm s}^2)\Psi' + \CH\Phi' + \left[2\CH' + (1+3c_{\rm s}^2)\CH^2\right]\Phi\\
	%&\qquad= a^2\frac{\kappa}{2}\left[c_{\rm s}^2\rho D_\rho + \delta p_{\not= \rm ad} + \frac{2}{3}\Delta\Pi\right] + K\Delta\CC + 3\CH c_{\rm s}^2 K \CC',\nonumber\\
	&\Psi - \Phi = a^2\kappa\Pi + 4K\CC, \label{eq::grav_slip}
\end{align}
\item for vector modes:
\begin{align}
	\left(\Delta - 2K\right)Q_i &= -a^2 2\kappa (\rho + p)(V_i - Q_i) + 4K\CC_i', \\
	Q_i' + 2\CH Q_i &= a^2\kappa\Pi_i + 4K\CC_i, \label{eq::vec_mod_2}
\end{align}
\item for tensor modes:
\begin{align}
	f_{ij}'' + 2\CH f_{ij}' + \left(6K - \Delta\right)f_{ij} = a^2\kappa \Pi_{ij}. \label{eq::H_ij_CC}
\end{align}
\end{enumerate}
The conservation of the topological term $\nabla_\nu\Top^{\nu\mu} = 0$ leads to
\begin{align}
	\CC'' + 2\CH\CC' - \left(\Delta + 4K\right)\CC &= \Phi-\Psi, \label{eq::bi-connection_CC_1} \\
	\CC_i'' + 2\CH\CC_i' - \left(\Delta + 2K\right)\CC_i &=  -\left(Q_i' + 2\CH Q_i\right). \label{eq::bi-connection_CC_2}
\end{align}
Combining these last two equations with, respectively, \eqref{eq::grav_slip} and~\eqref{eq::vec_mod_2}, we obtain
\begin{align}
	\CC'' + 2\CH\CC' - \Delta\CC &= -a^2\kappa \Pi, \label{eq::bi-connection_1_wave} \\
	\CC_i'' + 2\CH\CC_i' + \left(2K -\Delta\right) \CC_i &= -a^2\kappa \Pi_i. \label{eq::bi-connection_2_wave}
\end{align}
These are wave equations for $\CC$ and $\CC_i$ that are sourced, respectively, by the scalar and vector parts of the anisotropic stress. Finally, since the conservation equations~\eqref{eq::cons_1}--\eqref{eq::cons_V_mode} do not depend on the Einstein equation, they are unchanged in the bi-connection theory.

\subsection{Interpretation of $\CC$ and $\CC^i$}
\label{sec::interp_C}

Compared to the Einstein equation, the weak field limit of the bi-connection theory features two additional variables: the scalar mode $\CC$ and the vector mode $\CC^i$, which are constrained by the wave equations~\eqref{eq::bi-connection_1_wave} and \eqref{eq::bi-connection_2_wave}. These variables quantify the additional degrees of freedom appearing with the introduction of the reference curvature $\bar R_{\mu\nu}$. In particular, they can be related to properties of the reference observer induced by $\bar R_{\mu\nu}$, in the cases $K\not=0$. We recall that this observer is defined by a 4-velocity $G^\mu$ such that\footnote{While the property ${\rm dim}(\ker(\bar R_{\mu\nu})) = 1$ defines, up to a factor, a reference vector field $G^\mu$, we are unsure if this vector can be normalised  to be timelike with respect to the physical spacetime metric everywhere. However, the fact that it is possible at zeroth order, as $G^\mu \propto n^\nu$, suggests that it is not unphysical to consider that property to hold at first order. In any case, the weak field equations of Sec.~\ref{sec::gauge_eq_bi_metric} do not depend on the existence of such a normalisation.} $G^\nu \bar R_{\mu\nu} \coloneqq 0$ and $G^\mu G^\nu g_{\mu\nu} = -1$. We have
\begin{align}
	G^\mu	&= \frac{1}{a}\left( 1-\phi; -D^i \CX ' - {\CX^i}'  \right).
\end{align}
To have a better view of the link between $\CC$, $\CC_i$ and $G^\mu$, let us also introduce the vector normal to the foliation of constant time $n^\mu$, i.e. the foliation relative to the coordinates in which~\eqref{eq::phys_g} and \eqref{eq::eta_munu} hold, along with the 4-velocity $u^\mu$ of the fluid described by the energy-momentum tensor~{\eqref{eq::T^0_0}--\eqref{eq::T^i_j}}:
\begin{align}
	n^\mu	&\coloneqq \frac{1}{a}\left( 1-\phi; -D^i B + S^i \right), \\
	u^\mu	&\coloneqq \frac{1}{a}\left( 1-\phi; D^i v + v^i \right).
\end{align}
The tilt between these three vectors is
\begin{align}
	u^\mu - n^\mu	&=  \frac{1}{a}\left( 0; D^i(V - \sigma) + V^i - Q^i\right), \\
	n^\mu - G^\mu	&=  \frac{1}{a}\left( 0; D^i(\CC' + \sigma) + {\CC^i}' + Q^i\right), \\
	u^\mu - G^\mu	&=  \frac{1}{a}\left( 0; D^i(\CC' + V) + {\CC^i}' + V^i\right), \label{eq::tilt_u-G}
\end{align}
where we introduced the gauge invariant variable $V_i \coloneqq v_i + F_i'$. We see that $\CC$ and $\CC_i$ quantify part of the tilt between the constant time, the fluid and the reference observers. The 4-acceleration of the latter is given by
\begin{align}
	\tensor[^{G}]{a}{^\mu}	&\coloneqq G^\nu\nabla_\nu G^\mu \nonumber\\
						&= \frac{1}{a^2} \delta^\mu_i\left[D^i\left(\Phi - \CC'' - \CH\CC'\right) - \left({\CC^i}' + {Q^i}\right)'  -\CH \left({\CC^i}' + {Q^i}\right)\right], \label{eq::acc_G}
\end{align}
and its vorticity is
\begin{align}
	\Omega_{\mu\nu}		&\coloneqq b_{\alpha[\mu} b^{\beta}{}_{\nu]} \nabla_\beta G^\alpha \nonumber \\
						&= -a \delta^i_{[\mu} \delta^j_{\nu]}D_{i} \left(\CC_{j}' +Q_{j}\right), \label{eq::w_ij_G}
\end{align}
with $b_{\mu\nu} \coloneqq g_{\mu\nu} + G_\mu G_\nu$. We can also compute the curvature perturbation $\delta\CR^{(G)}$ of the reference observer. It is defined as the first order of the scalar spatial curvature in a gauge choice where the scalar part of the tilt between $n^\mu$ and $G^\mu$ is zero, i.e. $\sigma = - \CC'$. We have
\begin{align}
	\delta\CR^{(G)} = \frac{4}{a^2}\left(\Delta + 3K\right)\left(\Psi + \CH\CC'\right).
\end{align}
Therefore, $\Psi + \CH\CC'$ quantifies the curvature perturbation of the reference observer rest frames.\saut

As seem with Eq.~\eqref{eq:biCoEq}, the main difference between general relativity and the bi-connection theory is that matter only induces a departure of the physical curvature from the reference curvature. Therefore, it seems more relevant to consider not directly the perturbation of the spatial scalar curvature, i.e. $\delta\CR_{|_{\sigma = -\CC'}}$, but rather the perturbation of the spatial scalar curvature departure, i.e. $\delta\left(\CR - b^{\mu\nu}\bar R_{\mu\nu}\right)_{|_{\sigma = -\CC'}}$, which is
\begin{align}
	\delta\left(\CR - b^{\mu\nu}\bar R_{\mu\nu}\right)_{|_{\sigma = -\CC'}} = \frac{4}{a^2}\Delta\left(\Psi + \CH\CC' -K \CC\right). \label{eq:R_departure}
\end{align}

Therefore, Eqs.~\eqref{eq::tilt_u-G}, \eqref{eq::acc_G} and \eqref{eq:R_departure} suggest the introduction of the following gauge invariant variables:
\begin{enumerate}
	\item $\tilde\Phi \coloneqq \Phi - \CC'' - \CH\CC'$, i.e. the scalar mode of the acceleration of the reference observer,
	\item $\tilde\Psi	\coloneqq \Psi + \CH \CC' - K \CC$, i.e. the perturbation of the spatial scalar curvature departure of the reference observer rest frames,
    \item $\tilde\delta \coloneqq \delta - 3\CH(1 + w)\CC'$, i.e. the density perturbation in the reference observer rest frames,
	\item $\tilde V	\coloneqq V + \CC'$, i.e. the scalar mode of the tilt between the reference observer and the fluid,
	\item $\tilde V_i	 \coloneqq V_i + \CC_i'$, i.e. the vector mode of the tilt between the reference observer and the fluid,
    \item $\tilde Q_i	\coloneqq Q_i + \CC_i'$, quantifying the vorticity of the reference observer.
\end{enumerate}
What is remarkable is that by introducing these variables in the system~\eqref{eq::Poisson_CC}--\eqref{eq::vec_mod_2}, the scalar mode $\CC$, which is still sourced by $\Pi$ with Eq.~\eqref{eq::bi-connection_1_wave}, becomes the only source for the reference gravitational slip $\tilde\Psi - \tilde\Phi$, and disappears from the rest of the scalar mode equations
\begin{align}
	%&\Delta\tilde\Psi = a^2\frac{\kappa}{2}\rho D_\rho + \CH\left(\tilde\Psi' - \tilde\Phi'\right), \label{eq::Poisson_CC_tilde} \\
    &\Delta\tilde\Psi = a^2\frac{\kappa}{2}\rho\tilde\delta +   \CH\left(4\tilde\Psi' - \tilde\Phi' + 3\CH\tilde\Phi\right), \label{eq::Poisson_CC_tilde} \\
	&\tilde\Psi' + \CH \tilde\Phi = -a^2\frac{\kappa}{2}(\rho+p)\tilde V, \\
 	&\tilde\Psi'' + 2\CH\tilde\Psi' + \CH\tilde\Phi' + (2\CH' + \CH^2)\tilde\Phi = a^2\frac{\kappa}{2}\left(c_{\rm s}^2\rho\tilde\delta + \delta p_{\not= \rm ad} + \frac{2}{3}\Delta\Pi\right), \\
	%&\tilde\Psi'' + \CH(2 + 3c_{\rm s}^2)\tilde\Psi' + \CH\tilde\Phi' + \left[2\CH' + (1+3c_{\rm s}^2)\CH^2\right]\tilde\Phi \nonumber \\
	%		&\qquad = a^2\frac{\kappa}{2}\left[c_{\rm s}^2\rho D_\rho + \delta p_{\not= \rm ad} + \frac{2}{3}(\Delta+3K)\Pi\right],\nonumber\\
	&\tilde\Psi - \tilde\Phi = \left(\Delta + 3K\right)\CC. \label{eq::grav_slip_tilde}
\end{align}
The vector modes equations become
\begin{align}
	%&\tilde Q_i'' + 2\CH \tilde Q_i'+ 2\CH' \tilde Q_i - 2\left(\Delta - 2K\right)\tilde Q_i = a^24\kappa (\rho + p)\tilde V_i, \\
	&\left(\Delta - 2K\right)\tilde Q_i = -2a^2\kappa (\rho + p)\left(\tilde V_i - \tilde Q_i\right) + \left(\Delta + 2K\right){\CC'_i}, \\
	&\tilde Q_i' + 2\CH \tilde Q_i = \left(\Delta + 2K\right)\CC_i. \label{eq::vec_mod_2_tilde}
\end{align}
The conservation equations for matter become
\begin{align}
    &\tilde\delta' + 3\CH\left[\left(c_{\rm s}^2 - w\right)\tilde\delta + \delta p_{\not= \rm ad}/\rho\right] + (1+w)\left(\Delta\tilde V - 4\tilde\Psi' + \tilde\Phi'\right) = 0, \\
	%&D_\rho' - 3HwD_\rho = -(1+w)\Delta \tilde V + (\Delta + 3K)\left[ 2\CH \Pi/\rho + (1+w)\CC' \right], \\
    &\tilde V' + \CH\left(1-3c_{\rm s}^2\right)\tilde V + \tilde \Phi + \frac{1}{\rho + p}\left[c_{\rm s}^2\rho \tilde\delta + \delta p_{\not= \rm ad} + \frac{2}{3}\left(\Delta + 3K\right)\Pi\right] = 0.
	%&\tilde V' + \CH \tilde V + \tilde \Phi + \frac{1}{\rho + p}\left[c_{\rm s}^2\rho D_\rho + \delta p_{\not= \rm ad} + \frac{2}{3}\left(\Delta + 3K\right)\Pi\right] = 0.
\end{align}
In a non-interacting multi fluid approach, these last two equations are fulfilled for each fluid component.\saut

Unfortunately, {\it a priori}, $\CC$ and $\CC_i$ cannot be totally removed from the equations by a change of gauge invariant variables. Therefore, in the general case of our model, there necessarily are two additional variables with respect to the weak field limit of the Einstein equation.\saut

\begin{remark}
	The weak field equations~\eqref{eq::Poisson_CC}--\eqref{eq::vec_mod_2}, i.e. written as functions of $\Phi$, $\Psi$, $V$ and $V^i$, reduce to the standard weak field equations of general relativity in the case $K=0$. However, the tilde variables $\tilde\Phi$, $\tilde\Psi$, $\tilde V$, and $\tilde V^i$ do not reduce to the non tilde ones. This means that if the usual scale invariant initial conditions taken for $\Psi$ are shifted to $\tilde\Psi$, then, even the case $K=0$ can lead to a different prediction on the CMB power spectrum than from with the Standard Model. However, it is not clear to us if a proper justification of this change of initial conditions from $\Psi$ to $\tilde\Psi$ can be found.
\end{remark}

\begin{remark}{Reference \citep{2022_Vigneron_c} showed that $G^\mu$ corresponds to the 4-velocity of a Galilean observer, i.e. defining the Newtonian notion of inertial frames in the non-relativistic limit. Therefore, in the case $\CC = 0 = \CC^i$, the usual gauge invariant variables have an elegant interpretation:  $\Phi$ describes the acceleration of inertial frames, $\Psi$ their curvature perturbation, $Q^i$ their vorticity, and $V$ and $V^i$ the tilt of the fluid with respect to these frames.}
\end{remark}

\section{Blind curvature and cosmological data}
\label{sec:obs}

In this section we fit our model with CMB, BAO, and SN1a data. Throughout the section, the ``$0$'' subscript for current time values of the $\Omega$-cosmological parameters will be omitted.

\subsection{Methods}

We make use of a modified\footnote{Note that the correspondence between the Newtonian gauge variables used in {\sc class} (defined in \citet{1995_Ma_et_al}) and our notation is:
\begin{align}
    \Psi^{\rm Ma} = \Phi \quad ; \quad \Phi^{\rm Ma} = \Psi \quad ; \quad   \delta^{\rm Ma} = \delta \quad ; \quad \theta^{\rm Ma} = - k^2 V \quad ; \quad \sigma^{\rm Ma} = \frac{2}{3} \frac{k^2\Pi}{\rho + p},
\end{align}
where $k$ is the wave number of the harmonic decomposition.} version of the public {\sc class}\footnote{\url{https://lesgourg.github.io/class_public/class.html}} code \citep{2011_Blas_et_al} and  run Markov-chain Monte Carlo runs using the Metropolis-Hasting algorith implemented in {\sc Monte Python~v3}\footnote{\url{https://github.com/brinckmann/montepython_public}} \citep{2013_Audren_et_al,2019_Brickmann_et_al}.
We consider various combinations of the Planck TT/TE/EE and ``conservative'' lensing potential power spectra \cite{2020_Planck_VI}, measurements of the BAO from the CMASS and LOWZ galaxy samples of BOSS DR12 at $z = 0.38$, 0.51, and 0.61 \cite{2017_BOSS}, and the BAO measurements from 6dFGS at $z = 0.106$ and SDSS DR7 at $z = 0.15$  \cite{2011_Beutler_et_al,2015_Ross_et_al}; the Pantheon+ SNIa catalog compiles information about the luminosity distance to over 1600 SN1a  in the redshift range $0.01 < z < 2.3$ \citep{2022_Brout_et_al}. 
In all runs, we use large flat priors on $H_0$, the  baryon and cold dark matter energy density $\omega_{\rm b}$ and $\omega_{\rm cdm}$, respectively, and vary the curvature density fraction $\Omega_K\in[-0.5,0.5]$. When considering Planck, we also include the amplitude and tilt of the scalar perturbations $A_{\rm s}$ and $n_{\rm s}$, respectively (see next section for a proper definition), and the reionisation optical depth $\tau_{\rm reio}$. We model free-streaming neutrinos as two massless species and one massive with $m_\nu=0.06$ eV. We use {\sc Halofit} to estimate the non-linear matter clustering \cite{2003_Smith_et_al,2012_Takahashi_et_al}.
We consider chains to be converged using the conventional Gelman-Rubin criterion $|R -1|\lesssim0.01$ \citep{1992_Gelman_et_al} . 
To analyze the chains and produce our figures, we use {\sc GetDist} \cite{2019_Lewis}.

\subsection{Initial conditions}

%\subsection{Choice of $\CC$ and $\CC_i$}
\label{sec::choice_CC}

From the wave equations~\eqref{eq::bi-connection_1_wave} and \eqref{eq::bi-connection_2_wave}, we see that $\CC$ and $\CC^i$ are sourced by the scalar and vector parts of the fluid anisotropic stress. This leads to three possible situations:
\begin{enumerate}
	\item ($\Pi = 0$; $\Pi_i = 0$) and ($\CC = 0$; $\CC_i = 0$): This is the simplest case.  The cosmological model defined via this system along with the expansion laws~\eqref{eq::H2_bi-connection} and~\eqref{eq::ddota_bi-connection} has the same number of variables as the $\Lambda$CDM model with curvature, the only difference being the presence or not of the coupling terms with that curvature. 
	\item ($\Pi = 0$ and/or $\Pi_i = 0$) and ($\CC \not= 0$ and/or $\CC_i \not= 0$): Choosing $\CC = 0 = \CC_i$ without anisotropic stress is a restriction to the generality of the weak field equations. In particular, the gravitational slip, i.e. $\Psi - \Phi$, is not necessarily zero but sourced by~$\CC$. It is not clear to us if this choice is physical, especially since $\CC$ and $\CC_i$ vanish in the non-relativistic limit, as shown in Appendix~F of \citep{2022_Vigneron_c}. 
	\item ($\Pi \not= 0$; $\Pi_i \not= 0$) and ($\CC \not= 0$; $\CC_i \not= 0$): The presence of anisotropic stress necessarily implies the presence of $\CC$ and $\CC_i$, as shown by the wave equations~\eqref{eq::bi-connection_1_wave} and \eqref{eq::bi-connection_2_wave}, and therefore implies the presence of additional parameters with respect to the weak field equations of general relativity.
\end{enumerate}

Since anisotropic stress plays a non-negligible role in the CMB power spectrum due to the presence of free-streaming neutrinos (see e.g. \citep{2004_Bashinsky_et_al}), we will consider this third case when fitting Planck data with this cosmological model. Furthermore, with anisotropic stress being zero initially, we will consider a zero initial condition for $\CC$ and $\CC^i$. A proper justification for a more complex initial condition on these variables remains to be given, and is left for a future work.\saut

In the $\Lambda$CDM model, the parametrisations of the primordial power spectrum for non flat cases is debated (e.g. \citep[][]{1995_White_et_al,2003_Efstathiou, 2019_Handley,2021_Thavanesan_et_al}), essentially because there is no consensus on a non-flat inflationary scenario. Given the lack of such a scenario in the context of our model, the issue remains. Therefore, we assume for simplicity the standard parametrisation used by {\sc class}:
\begin{align}
    \Delta(k) = A_{\rm s} \left(\frac{k}{k_\star}\right)^{n_{\rm s}-1}, 
\end{align}
with $k_\star=0.05h/$Mpc$^{-1}$ the conventional pivot scale. As mentioned previously, we vary $A_{\rm s}$ and $n_{\rm s}$ within broad flat priors in analyses that include Planck data.

\subsection{Results}

\begin{figure}[t]
	\centering
	\includegraphics[width=12cm]{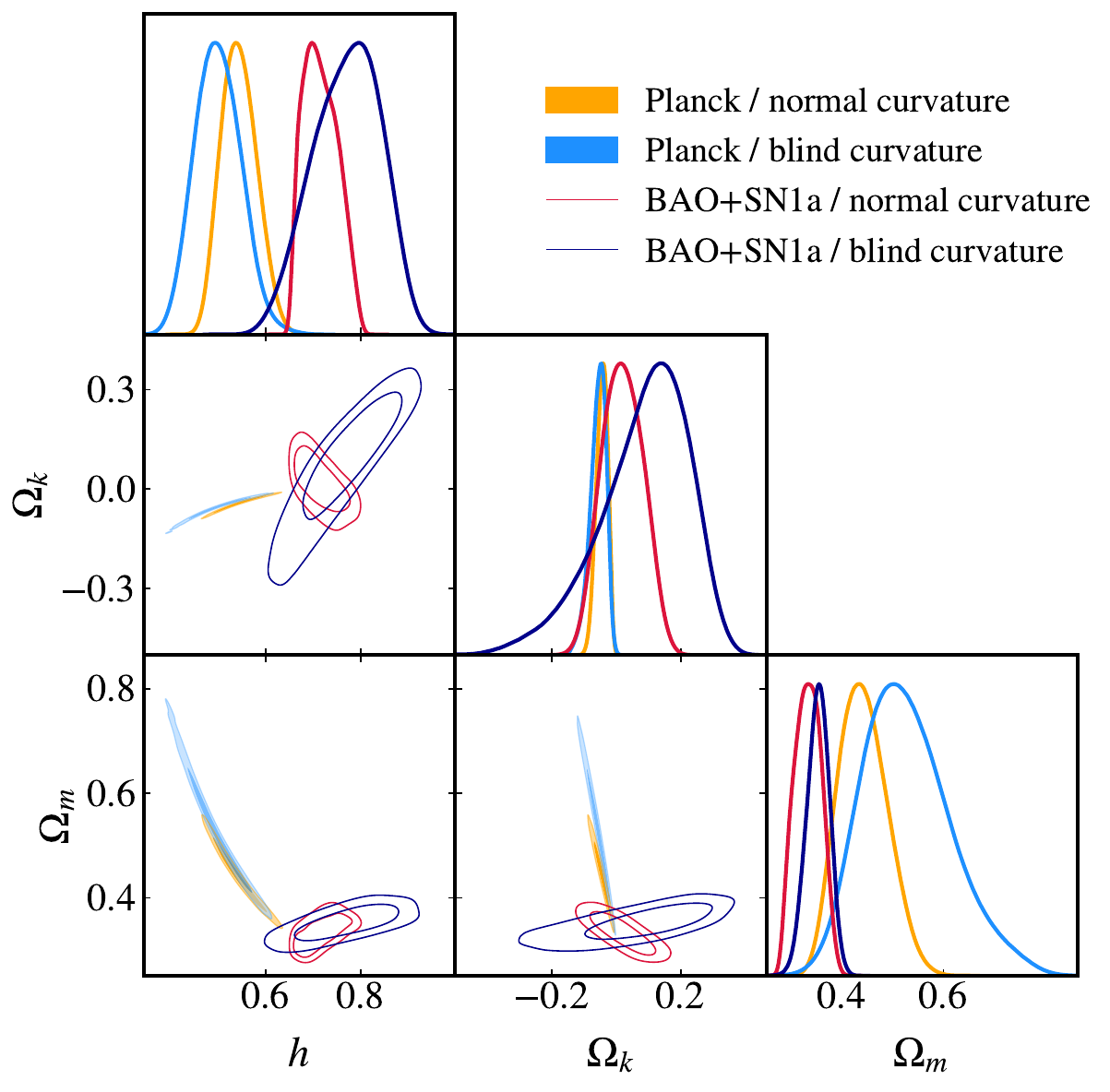}
	\caption{\label{fig:1} 2D posterior distributions of $\{h,\Omega_K,\Omega_{\rm m}\}$ in the standard case (red/orange) and in the blind curvature (blue) model. We compare constraints from Planck (filled) alone, to that obtained from BAO+SN1a (empty).}
\end{figure}

\begin{figure}[t]
	\centering
	\includegraphics[width=12cm]{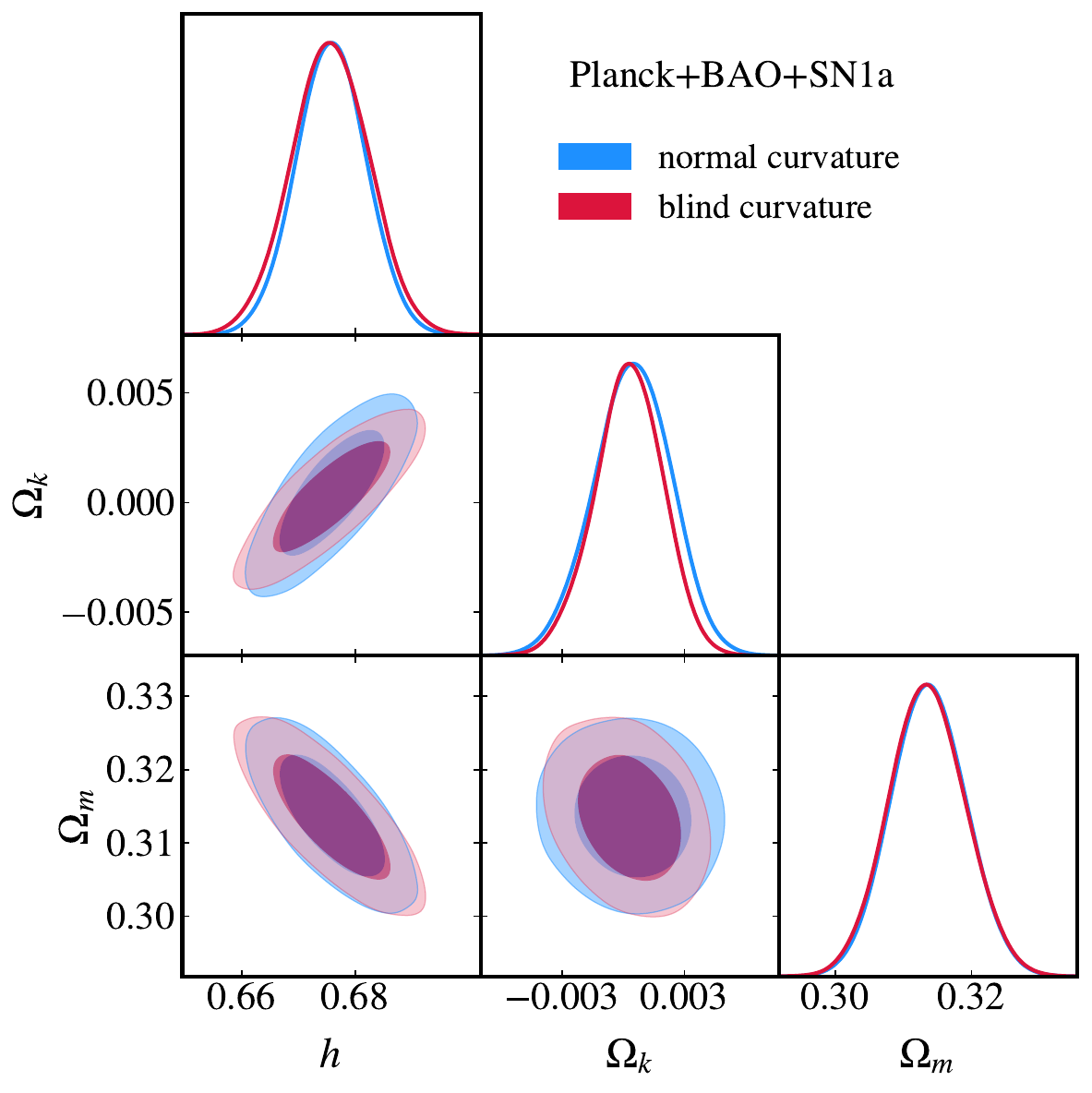}
	\caption{ \label{fig:2} 2D posterior distributions of $\{h,\Omega_K,\Omega_{\rm m}\}$ in the standard case and in the blind curvature model for the combined analysis Planck+BAO+SN1a.}
\end{figure}

\begin{table}[t]
	\centering
\scalebox{0.8}{\begin{tabular}{|c|c|c|c|c|}
 \hline
 \hline

 & \multicolumn{2}{c|}{Planck} &  \multicolumn{2}{c|}{Planck+BAO+SN1a} \\
 \hline
curvature? &  normal &  blind &  normal &  blind \\
 \hline

$h$
	 & $0.543(0.544)^{+0.035}_{-0.041}$ 
	 & $0.499(0.466)\pm 0.048$ 
	 & $0.6758(0.6738)\pm 0.0062$ 
	 & $0.6756(0.6744)\pm 0.0069$ 
	 \\
$\Omega_{K }$
	 & $-0.044(-0.042)^{+0.020}_{-0.015}$ 
	 & $-0.057(-0.072)^{+0.029}_{-0.019}$ 
	 & $0.0005(0.0002)\pm 0.0019$ 
	 & $0.0003(-0.0001)^{+0.0017}_{-0.0016}$ 
	 \\
$\Omega_{\rm m}$
	 & $0.442(0.563)^{+0.043}_{-0.053}$ 
	 & $0.528(0.424)^{+0.071}_{-0.099}$ 
	 & $0.3137(0.3158)\pm 0.0055$ 
	 & $0.3135(0.3143)\pm 0.0056$ 
	 \\
   \hline

$\omega_{\rm cdm }$
	 & $0.1182(0.1183)\pm 0.0015$ 
	 & $0.1182(0.1169)\pm 0.0015$ 
	 & $0.1200(0.1202)\pm 0.0013$ 
	 & $0.1199(0.1199)\pm 0.0013$ 
	 \\
$10^{2}\omega_{\rm b }$
	 & $2.258(2.258)\pm 0.017$ 
	 & $2.257(2.274)\pm 0.017$ 
	 & $2.235(2.243)\pm 0.015$ 
	 & $2.235(2.245)\pm 0.015$ 
	 \\
$10^{9}A_{\rm s }$
	 & $2.067(2.089)\pm 0.035$ 
	 & $2.065(2.048)\pm 0.035$ 
	 & $2.101(2.089)\pm 0.030$ 
	 & $2.101(2.109)\pm 0.029$ 
	 \\
$n_{\rm s }$
	 & $0.9701(0.9702)\pm 0.0048$ 
	 & $0.9698(0.9728)\pm 0.0047$ 
	 & $0.9643(0.9636)\pm 0.0044$ 
	 & $0.9646(0.9649)\pm 0.0043$ 
	 \\
$\tau_{\rm reio }$
	 & $0.0486(0.0539)\pm 0.0083$ 
	 & $0.0482(0.047)\pm 0.0081$ 
	 & $0.0545(0.0507)\pm 0.0072$ 
	 & $0.0546(-0.0001)\pm 0.0072$ \\

\hline\hline
 \end{tabular}}
	\caption{\label{tab1} Mean (best fit) $\pm 1\sigma$ errors in the $\Lambda$CDM+$\Omega_k$ model with normal or blind curvature, reconstructed from either Planck or Planck+BAO+SN1a.  }

\end{table}

\begin{table}[t]
	\centering
\scalebox{0.8}{\begin{tabular}{|c|c|c|}
 \hline
 \hline

 & \multicolumn{2}{c|}{BAO+SN1a} \\
 \hline
curvature? &  normal &  blind  \\
 \hline

$h$
	 & $0.715(0.697)_{-0.046}^{+0.03}$
	 & $0.770(0.851)_{-0.071}^{+0.079}$  

	 \\
$\Omega_{K }$
	 &  $0.0199(0.085)_{-0.068}^{+0.067}$
	 & $0.086(0.150)_{-0.1}^{+0.17}$

	 \\
$\Omega_{\rm m}$
	 & $0.353(0.389)_{-0.049}^{+0.046}$
	 & $0.406(0.457)_{-0.073}^{+0.13}$

	 \\

\hline\hline
 \end{tabular}}
	\caption{\label{tab2} Mean (best fit) $\pm 1\sigma$ errors in the $\Lambda$CDM+$\Omega_k$ model with normal or blind curvature, reconstructed from BAO+SN1a.  }

\end{table}

We perform three sets of analyses: Planck TT/TE/EE  (no lensing),  BAO+SN1a on their own, and finally the combination of Planck TT/TE/EE+lensing+BAO+SN1a. 
For each of these combinations of datasets, the fits are performed with the $\Lambda$CDM model (dubbed ``normal curvature'') for comparison and with our model (dubbed ``blind curvature''). We provide the mean (best fit) $\pm1\sigma$ error reconstructed for each parameter in Tables~\ref{tab1} and \ref{tab2}.\saut

Figure~\ref{fig:1} presents the posterior distributions of the Planck alone (filled) and BAO+SN1a (empty) fits, for $\Lambda$CDM (orange/red) and the blind model (light/dark blue).  The main difference brought by the blind model is an increase of the uncertainty, while the average values of the parameters remain approximately the same. In particular, the curvature tension mentioned earlier between the Planck and the BAO+SN1a datasets is still present in the blind model, even slightly enhanced.
Interestingly, the degeneracy directions between $H_0$ and $\Omega_{K}$ in the BAO+SN1a analysis are opposite between the two models. This can be understood from the first order in $\Omega_{K}$ of the angular diameter distance formula~\eqref{eq:dA} (which is the main formula governing the BAO+SN1a fit), assuming $\Omega_{\rm m}$ and $\Omega_{\Lambda}$ as fixed variables independent of $H_0$ and $\Omega_{K}$, which leads to:
\begin{align}
    \Omega_{K}^{\Lambda \rm CDM} &= -\frac{6\CI_1(z)}{\CI_1^3(z) - 6\CI_2(z)} + \frac{6(1+z)d_{\rm A}(z)}{\CI_1^3(z) - 6\CI_2(z)}\, H_0 +\bigO{\Omega_{K}^2}, \\
     \Omega_{K}^{ \rm blind} &= -\frac{6}{\CI_1^2(z)} + \frac{6(1+z)d_{\rm A}(z)}{\CI_1^3(z)}\, H_0 +\bigO{\Omega_{K}^2},
    %d^{\Lambda{\rm CDM}}_{\rm A}(z) &= \frac{1}{H_0}\left(\frac{\mathcal{I}_1(z)}{(1+z)} + \frac{\mathcal{I}_1^3(z)-6\mathcal{I}_2(z)}{6(1+z)}\,\Omega_K\right) +\bigO{\Omega_K^2}, \\
    %d^{\rm blind}_{\rm A}(z) &= \frac{1}{H_0}\left(\frac{\mathcal{I}_1(z)}{(1+z)} + \frac{\mathcal{I}_1^3(z)}{6(1+z)}\,\Omega_K\right) +\bigO{\Omega_K^2},
\end{align}
where
\begin{align}
    \mathcal{I}_1(z) &\coloneqq \int_0^z \frac{1}{\left[\Omega_{\rm m}(1+y)^3 + \Omega_{\Lambda}\right]^{1/2}}\dd y > 0, \\
    \mathcal{I}_2(z) &\coloneqq \int_0^z\frac{(z+1)^2/2}{\left[\Omega_{\rm m}(1+y)^3 + \Omega_{\Lambda}\right]^{3/2}}\dd y > 0.
\end{align}
In our model,  the degeneracy (of $\Omega_{K}$ as a function of $H_0$) has a positive slope with factor $\frac{6(1+z)d_{\rm A}(z)}{\CI_1^3(z)}$. In the $\Lambda$CDM model, this slope is $\frac{6(1+z)d_{\rm A}(z)}{\CI_1^3(z) - 6\CI_2(z)}$ and can be negative if ${\mathcal{I}_1^3(z) < 6\,\mathcal{I}_2(z)}$, as is the case with BAO+SN1a data. The slope of the degeneracy depends on the redshift, which explains why it is different between the Planck data and the BAO+SN1a datasets.\saut

Figure~\ref{fig:2} presents the posterior distributions for the combined Planck+BAO+SN1a fit in the $\Lambda$CDM ``normal curvature'' model (red) and the blind model (blue). No significant difference can be found between the two models. In particular, the curvature is still tightly constrained around zero.\saut

In all the fits, we find that the additional gauge invariant variables present in our model at the level of perturbations play a negligible role, the difference with $\Lambda$CDM coming mainly from the modified background expansion laws~\eqref{eq::Omega_law}.\saut

Overall, the difference with respect to the best fit with the $\Lambda$CDM model and our model is not significant. In particular, while we still have a preference for a spherical universe when Planck data alone are used, the increase in the Hubble tension is still present. With respect to the discussion in Sec.~\ref{sec:blind}, this shows that the main constraints given by cosmological data on the value of the curvature parameter come from the geometrical effects of that curvature.

\section{Conclusion}
\label{sec::disc}

We derived the homogeneous and isotropic solution of the bi-connection theory developed in~\citep{2022_Vigneron_c}, in which a term related to topology is added in the Einstein equation. The new expansion laws do not feature the curvature parameter anymore, regardless of its value, i.e. $\Omega_{\not= K} = 1, \ \forall \, \Omega_K$. In other words, in this cosmological model, the expansion scenario is equivalent for a Euclidean, spherical, or hyperbolic universe: i.e. expansion is blind to the spatial curvature. The first order perturbations around this homogeneous solution features two new gauge invariant variables compared to the Standard Model. A scalar and a vector mode, both sourced by the anisotropic stress of the fluid, and which can be related to a reference observer.\saut

We tested our model against observations. In the different combinations of datasets, no significant difference was found between our model and the $\Lambda$CDM model. In particular, the curvature tension between the Planck and the BAO datasets remains present within our model, and is even slightly increased, with Planck preferring a closed universe with $\Omega_{K,0} = -0.057 \pm 0.025$ and $H_0 = 50 \pm 5$ km/s/Mpc.\saut

Overall, these results show that removing the curvature parameter from the expansion laws does not significantly change its estimation from current cosmological data. Since the presence or not of spatial curvature is the main difference between our cosmological model and the $\Lambda$CDM model, only a better precision of the measurement of that parameter might enable us to distinguish between these two models, and therefore distinguish between general relativity and the bi-connection theory of~\citep{2022_Vigneron_c}. The study of a non-flat inflationary scenario under the framework of this theory is also an interesting perspective, especially since the flatness problem is not present with the blind expansion law anymore.

\section*{Acknowledgements}
Q.V. is supported by the Centre of Excellence in Astrophysics and Astrochemistry of Nicolaus Copernicus University in Toru\'n, and by the Polish National Science Centre under Grant No. 2022/44/C/ST9/00078. V.P. has received support from the European Research Council under the
European Union’s HORIZON-ERC-2022 (Grant Agreement No. 101076865).
This project has received support from the European Union’s Horizon 2020 research and innovation program under the Marie Skodowska-Curie Grant Agreement No. 860881-HIDDeN.
These results have been made possible thanks to LUPM's cloud computing infrastructure founded by Labex Ocevu, and France-Grilles.
We are grateful to Pierre Mourier for constant discussions and commentaries on the manuscript. We thank Julien Bel and Christian Marinoni for insightful discussions.

\appendix

\section{No effective spatial curvature in the Newtonian expansion law}
\label{eq::Exp_NEN}

\subsection{Motivation}

In Newtonian gravitation, expansion is constrained by the averaged second Friedmann equation
\begin{align}
		3\ddot a/a	= -\frac{\kappa}{2}\Saverage{\rho} + \Lambda - \Xi_{cd}\Xi^{cd}, \label{eq::exp_NR}
\end{align}
where $\Saverage{\cdot}$ is the spatial average over the whole (spatial) volume of $\Sigma$, and $\Xi_{cd}$ is traceless-transverse and represents anisotropic expansion. We will not consider $\Xi_{ij}$ further. That law is valid both for Newtonian gravitation on a Euclidean topology \citep{1955_Heckmann_et_al, 1997_Buchert_et_al, 2021_Vigneron} or for Newtonian gravitation on a non-Euclidean topology \citep{2022_Vigneron_b} and consequently was shown to derive from the non-relativistic limit of, respectively, general relativity and the bi-connection theory.\saut

However, while the second Friedman equation is explicitly obtained from the equations of Newtonian gravitation, the first Friedmann equation (i.e. featuring $H^2$) is only retrieved after integrating the former equation, as is well known in Newtonian cosmology, leading to
\begin{equation}
		3H^2	 = \kappa\Saverage{\rho} + \Lambda + C/a^2, \label{eq::H2_eff}
\end{equation}
where $C$ is an integration constant mimicking a spatial curvature term. To our knowledge, textbooks and references talking about Newtonian cosmology always assume that $C$ is free (even though it is generally taken to be zero).\saut

Since the main result of the present paper is the fact that spatial curvature should not be present anymore in the expansion law once we consider a theory compatible with the non-relativistic limit in any topology, then there seems to be a contradiction with~\eqref{eq::H2_eff}. The goal of this section is to show that this is not the case. We will show that a ``hidden'' condition can be found from the first order in the non-relativistic limit of either the Einstein equation or the bi-connection theory, which will constrain the integration constant $C$ to be zero (in either the Euclidean, spherical, or hyperbolic cases), thus retrieving the law~\eqref{eq::H2_bi-connection}. In other words, the expansion laws of non-relativistic (i.e. Newtonian) gravitation, if required to be compatible with either general relativity (Euclidean case) or the bi-connection theory (non-Euclidean case) must not feature an effective spatial curvature term.

\subsection{Derivation}

The derivation of that result requires the non-relativistic limit based on Galilean invariance that was developed by~\citep{1976_Kunzle}. We will not reintroduce this limit in the present paper. Rather, we will directly use some formulas obtained in \citep{2022_Vigneron_c}, which were derived from this limit. The Newtonian expansion law~\eqref{eq::exp_NR} corresponds to the volume average of the zeroth order (in~$1/c^2$) of the time-time components of the Einstein/bi-connection equations~\eqref{eq:biCoEq} (in other words, the average of the zeroth order of the Raychaudhuri equation). The first Friedmann law is obtained with the volume average of the first order of the spatial trace of~\eqref{eq:biCoEq} (in other words, the average of the first order of the trace of the 3+1-Ricci equation). That first order equation is given by the trace of Eq.~(135) in Appendix~E. of \citep{2022_Vigneron_c}, which gives
\begin{align}
3\left(\dot H + 3 H^2\right) + D_i P^i = 3\left(\frac{\kappa}{2}\rho + \Lambda\right), \label{zizi}
\end{align}
where $P^i$ is a vector depending on (post)-Newtonian terms that we do not need to detail, and $D_i$ is the spatial connection related to a constant curvature metric equivalent to $h_{ij}$ of Sec.~\ref{sec::Exp_law}.\saut

Equation~\eqref{zizi} is valid for Newtonian gravitation in either a Euclidean topology (i.e. from the non-relativistic limit of the Einstein equation) or in a spherical/hyperbolic topology (i.e. from the non-relativistic limit of the bi-connection theory). As a local equation, it does not constrain the Newtonian dynamics more even if the density is present, i.e. it does not need to be considered on top of the Poisson equation. Rather, it is a dictionary to calculate $P^i$, which can be related to the first order of the spatial curvature tensor~\citep{2022_Vigneron_c}. However, because the divergence $D_i P^i$ vanishes with an averaging procedure, the global property of this equation gives additional constraints on the global dynamics. Taking the spatial average of~\eqref{zizi} and using the acceleration law~\eqref{eq::exp_NR}, we obtain
\begin{equation}
	3H^2 = \kappa\Saverage{\rho} + \Lambda. \label{eq::h2_<>}
\end{equation}
This is the first Friedmann equation without a curvature term or integration constant, and, again, is obtain for Newtonian gravitation in any topology. This shows that, as for the homogenous solution of the bi-connection theory, the expansion law of any inhomogeneous (non-linear) solution of Newtonian gravitation is also blind to the spatial curvature.% in both of these theories, we have no freedom in choosing the integration constant $C$ when integrating~\eqref{eq::H2_eff} in the NR limit: $C=0$.

\IfFileExists{QV_mnras.bst}{\bibliographystyle{QV_mnras}}{\bibliographystyle{/Users/quentinvigneron/Documents/Travail/Research/tex_/QV_mnras}}
\IfFileExists{bib_General_blind.bib}{\bibliography{bib_General_blind}}{\bibliography{/Users/quentinvigneron/Documents/Travail/Research/tex_/bib_General_blind}}

\end{document}